\documentclass{llncs}
\usepackage[latin1]{inputenc}
\usepackage{longtable}
\usepackage[pdftex]{graphicx}
\DeclareGraphicsExtensions{.jpg,.png,.pdf}

\newcommand{\insertfigure}[5]{ 
\begin{figure}[#5]
\centering
\noindent\includegraphics[#4]{#1}
\caption{#2}
\label{#3}
\end{figure}}

\newcommand{\ie}{i.\,e.\ }
\newcommand{\eg}{e.\,g.\ }

\newcommand{\impact}[5]{\textsf{\small [#1$\,\vert\,$\uppercase{{\scriptsize#2}}] 
$\stackrel{#5}{\longrightarrow}$ [#3$\,\vert$\,\uppercase{{\scriptsize#4}}]} }

\newcommand{\impactn}[4]{\impact{#1}{#2}{#3}{#4}{-}}

\begin{document}
\frontmatter          
\pagestyle{headings}  
\mainmatter              
\title{A Comprehensive Model of Usability}
\titlerunning{A Comprehensive Model of Usability}  
%
\author{Sebastian Winter \and Stefan Wagner \and Florian Deissenboeck} 
\authorrunning{S. Winter \and S. Wagner \and F. Deissenboeck}   
%
%
\institute{Institut f\"ur Informatik \\ 
  Technische Universit\"{a}t M\"{u}nchen \\
  Boltzmannstr. 3, 85748 Garching b. M\"{u}nchen, Germany\\
\email{\{winterse, wagnerst, deissenb\}@in.tum.de}
}

\maketitle              

\begin{abstract}
Usability is a key quality attribute of successful software systems.
Unfortunately, there is no common understanding of the factors influencing usability and their interrelations.
Hence, the lack of a comprehensive basis for designing, analyzing and improving user interfaces.
This paper proposes a 2-dimensional model of usability that associates system properties with the activities carried out by the user.
By separating activities and properties, sound quality criteria can be identified, thus facilitating statements concerning their interdependencies.
This model is based on a tested quality meta-model that fosters preciseness and completeness.
A case study demonstrates the manner by which such a model aids in revealing contradictions and omissions in existing usability standards.
Furthermore, the model serves as a central and structured knowledge base for the entire quality assurance process, \eg the automatic generation of guideline documents.
\end{abstract}


\section{Introduction}

There is a variety of standards concerning the quality attribute \emph{usability} or \emph{quality in use}~\cite{bevan_2001_standards,seffah:2004}.
Although in general all these standards point in the same direction, due to different intuitive understandings of usability, they render it difficult to analyze, measure, and improve the usability of a system.
A similar situation also exists for other quality attributes, \eg \emph{reliability} or \emph{maintainability}.
One possibility to address this problem is to build a comprehensive model of the quality attribute.
Most models take recourse to the decomposition of quality proposed by Boehm et~al.~\cite{boehm:1978}.
However, this decomposition is still too abstract and imprecise to be used concretely for analysis and measurement.

More comprehensive models have been proposed for product quality in general~\cite{dromey:1995} or even usability~\cite{seffah:2006}.
However, these models have three problems:
First, they do not decompose the attributes and criteria to a level that is suitable for actually assessing them for a system.
Secondly, these models tend to omit rationale of the required properties of the system.
Thirdly, the dimensions used in these models are heterogeneous, \eg the criteria mix properties of the system with properties of the user.
The first problem constrains the use of these models as the basis for analyses.
The second one makes it difficult to describe impacts precisely and therefore to convince developers to use it.
The third problem hampers the revelation of omissions and inconsistencies in these models.
The approach to quality modeling by Broy, Deissenboeck, and Pizka~\cite{deissenb:2006} is one way to deal with these problems.
Using an explicit meta-model, it decomposes quality into system properties and their impact on activities carried out by the user.
This facilitates a more structured and uniform means of modeling quality.

\paragraph{Problem.}
Although usability is a key quality attribute in modern software systems, the general understanding of its governing factors is still not good enough for profound analysis and improvement.
Moreover, currently there is no objective criteria for evaluating usability.

\paragraph{Contribution.}
This paper proposes a comprehensive 2-dimensional model of usability based on a quality meta-model that facilitates a structured decomposition of usability and descriptions of the impacts of various facts of the system.
This kind of model has proven to be useful for the quality attribute \emph{maintainability}~\cite{deissenb:2006}.
Several benefits can be derived by using this type of model:

\begin{enumerate}
  \item The ability to reveal omissions and contradictions in current models and guidelines.
  \item The ability to generate guidelines for specific tasks automatically.
  \item A basis for (automatic) analysis and measurement.
  \item The provision of an interface with other quality models and quality attributes.
\end{enumerate}

We demonstrate the applicability of the 2-dimensional model in a case study of the ISO~15005~\cite{iso15005:2002} which involves domain-specific refinements.
By means of this model we are able to identify several omissions in the standard and suggest improvements.

\paragraph{Consequences.}
Based on the fact that we can pinpoint omissions and inconsistencies in existing quality models and guidelines, it seems advisable to use an explicit meta-model for usability models, precisely to avoid the weaknesses of the other approaches.
Furthermore, it helps to identify homogeneous dimensions for the usability modeling.
We believe that our model of usability is a suitable basis for domain- or company-specific models that must be structured and consistent.

\paragraph{Outline.}

In Sec.~\ref{sec:related} we describe prior work in the area of quality models for usability and the advances and shortcomings it represents.
In Sec.~\ref{sec:qmm}, using an explicit meta-model, we discuss the quality modeling approach.
The 2-dimensional model of usability that we constructed using this approach is presented in Sec.~\ref{sec:general_model}.
This model is refined to a specific model based on an ISO standard in the case study of Sec.~\ref{sec:case_study}.
The approach and the case study are discussed in Sec.~\ref{sec:discussion}.
In Sec.~\ref{sec:conclusions} we present our final conclusions.

\section{Related Work}
\label{sec:related}

This section describes work in the area of quality models for usability.
We discuss general quality models, principles and guidelines, and first
attempts to consolidate the quality models.

\subsection{Quality Models for Usability}

Hierarchical structures as quality models which focus mainly on \emph{quality assurance} have been developed.
A model first used by Boehm~\cite{boehm:1978} and McCall et~al.~\cite{mccall:1978} consists of three layers: factors, criteria, and metrics.
Consequently, the approach is referred to as the factor-criteria-metrics model (FCM model).
The high-level factors model the main quality goals.
These factors are divided into criteria and sub-criteria.
When a criterion has not been divided, a metric is defined to measure the criteria.
However, this kind of decomposition is too abstract and imprecise to be used for analysis and measurement.
In addition, since usability is not a part of the main focus, this factor is not discussed in detail.

In order to provide means for the operational measurement of usability several attempts have been made in the domain \emph{human-computer interaction} (HCI).
Prominent examples are the models from Shackel and Richardson~\cite{shackel:1991} or Nielsen~\cite{nielsen:1993}.
Nielsen~\cite{nielsen:1993}, for example, understands usability as a property with several dimensions, each consisting of different components.
He uses five factors: \emph{learnability}, \emph{efficiency}, \emph{memorability}, \emph{errors}, and \emph{satisfaction}.
\emph{Learnability} expresses how well a novice user can use the system, while the efficient use of the system by an expert is expressed by \emph{efficiency}.
If the system is used occasionally the factor \emph{memorability} is used.
This factor differentiates itself from \emph{learnability} by the fact that the user has understood the system previously.
Nielsen also mentions that the different factors can conflict with each other.

The ISO has published a number of standards which contain usability models for the operational evaluation of usability.
The ISO~9126-1~\cite{iso9126-1:2001} model consists of two parts.
The first part models the \emph{internal} as well as the \emph{external} quality, the second part the \emph{quality in use}.
The first part describes six characteristics which are further divided into sub-characteristics.
These measurable attributes can be observed during the use of the product.
The second part describes attributes for \emph{quality in use}.
These attributes are influenced by all six product characteristics.
Metrics are given for the assessment of the sub-characteristics.
It is important to note that the standard does not look beyond the sub-characteristics intentionally.

The ISO~9241 describes human-factor requirements for the use of software systems with user interface.
The ISO~9241-11~\cite{iso9241-11:1998} provides a framework for the evaluation of a running software system.
The framework includes the context of use and describes three basic dimensions of usability:
\emph{efficiency}, \emph{effectiveness}, and \emph{satisfaction}.

\subsection{Principles and Guidelines}

In addition to the models which define usability operationally, a lot of design principles have been developed.
Usability principles are derived from knowledge of the HCI domain and serve as a design aid for the designer.
For example, the ``eight golden rules of dialogue design'' from Shneiderman~\cite{shneiderman:1998} propose rules that have a positive effect on usability.
One of the rules, namely \emph{strive for consistency}, has been criticized by Grudin~\cite{grudin:1989} for its abstractness.
Grudin shows that consistency can be decomposed into three parts that also can be in conflict with each other.
Although Grudin does not offer an alternative model, he points out the limitations of the design guidelines.

Dix et~al.~\cite{dix:1998} argue as well that if principles are defined in an abstract and general manner, they do not help the designer.
In order to provide a structure for a comprehensive catalogue of usability principles Dix et~al. divide the factors which support the usability of a system into three categories:
\emph{learnability}, \emph{flexibility}, and \emph{robustness}.
Each category is further divided into sub-factors.
The ISO~9241-110~\cite{iso9241-110:2006} takes a similar approach and describes seven high-level principles for the design of dialogues:
\emph{suitability for the task}, \emph{self-descriptiveness}, \emph{controllability}, \emph{conformity with user expectations}, \emph{error tolerance}, \emph{suitability for individualization}, and \emph{suitability for learning}.
These principles are not independent of each other and some principles have an impact on other principles.
For example \emph{self-descriptiveness} influences \emph{suitability for learning}.
Some principles have a part-of relation to other principles.
For example, \emph{suitability for individualization} is a part of \emph{controllability}.
The standard does not discuss the relations between the principles and gives little information on how the principles are related to the overall framework given in~\cite{iso9241-11:1998}.

\subsection{Consolidated Quality Models for Usability}

There are approaches which aim to consolidate the different models.
Seffah et~al.~\cite{seffah:2006} applied the FCM model to the quality attribute \emph{usability}.
The developed model contains 10~factors which are subdivided into 26~criteria.
For the measurement of the criteria the model provides 127~metrics.

The motivation behind this model is the high abstraction and lack of aids for the interpretation of metrics in the existing hierarchically-based models.
Put somewhat differently, the description of the relation between metrics and high-level factors is missing.
In addition, the relation between factors, \eg \emph{learnability} vs. \emph{understandability}, are not described in the existing models.
Seffah et~al. also criticize the difficulty in determining how factors relate to each other, if a project uses different models.
This complicates the selection of factors for defining high-level management goals.
Therefore, in~\cite{seffah:2006} a consolidated model that is called \emph{quality in use integrated measurement model} (\textsc{Quim} model) is developed.

Since the FCM decomposition doesn't provide any means for precise structuring, the factors used in the \textsc{Quim} model are not independent.
For example, \emph{learnability} can be expressed with the factors \emph{efficiency} and \emph{effectiveness}~\cite{iso9241-11:1998}.

The same problem arises with the criteria in the level below the factors:
They contain attributes as well as principles, \eg \emph{minimal memory load}, which is a principle, and \emph{consistency} which is an attribute.
They contain attributes about the user (\emph{likeability}) as well as attributes about the product (\emph{attractiveness}).
And lastly, they contain attributes that are similar, \eg \emph{appropriateness} and \emph{consistency}, both of which are defined in the paper as capable of indicating whether visual metaphors are meaningful or not.

To describe how the architecture of a software system influences usability, Folmer and Bosch~\cite{folmer:2002} developed a framework to model the quality attributes related to usability.
The framework is structured in four layers.
The high-level layer contains \emph{usability definitions}, \ie common factors like \emph{efficiency}.
The second layer describes concrete measurable \emph{indicators} which are related to the high-level factors.
Examples of indicators are \emph{time to learn}, \emph{speed}, or \emph{errors}.
The third layer consists of \emph{usability properties} which are higher level concepts derived from design principles like \emph{provide feedback}.
The lowest layer describes the \emph{design knowledge} in the community.
Design heuristics, \eg the \emph{undo pattern}, are mapped to the \emph{usability properties}.
Van Welie~\cite{welie:1999} also approaches the problem by means of a layered model.
The main difficulty with layered models is the loss of the exact impact to the element on the high-level layer at the general principle level when a design property is first mapped to a general principle.

Based on Norman's action model~\cite{norman:1986} Andre et~al. developed the \textsc{User Action Framework}~\cite{andre:2001}.
This framework aims toward a structured knowledge base of usability concepts which provides a means to classify, document, and report usability problems.
By contrast, our approach models system properties and their impact on activities.

\subsection{Summary}
As pointed out, existing quality models generally suffer from one or more of the following shortcomings:

\begin{enumerate}

\item \emph{Assessability.}
Most quality models contain a number of criteria that are too coarse-grained to be assessed directly.
An example is the \emph{attractiveness} criterion defined by the ISO~9126-1~\cite{iso9126-1:2001}.
Although there might be some intuitive understanding of attractiveness, this model clearly lacks a precise definition and hence a means to assess it.

\item \emph{Justification.}
Additionally, most existing quality models fail to give a detailed account of the impact that specific criteria (or metrics) have on the user interaction.
Again the ISO standard cited above is a good example for this problem, since it does not provide any explanation for the presented metrics.
Although consolidated models advance on this by providing a more detailed presentation of the relations between criteria and factors, they still lack the desired degree of detail.
An example is the relationship between the criterion \emph{feedback} and the factor \emph{universality} presented in~\cite{seffah:2006}.
Although these two items are certainly related, the precise nature of the relation is unclear.

\item \emph{Homogeneity.}
Due to a lack of clear separation of different
aspect of quality most existing models exhibit inhomogeneous sets of quality criteria.
An example is the set of criteria presented in~\cite{seffah:2006} as it mixes attributes like \emph{consistency} with mechanisms like \emph{feedback} and principles like \emph{minimum memory load}.

\end{enumerate}

\section{A 2-Dimensional Approach to Model Quality}
\label{sec:qmm}

To address the problems with those quality models described in the previous section we developed the novel two-dimensional quality meta-model \textsc{Qmm}.
This meta-model was originally based on our experience with modeling maintainability~\cite{deissenb:2006}, but now also serves as a formal specification for quality models covering different quality attributes like \emph{usability} and \emph{reliability}.
By using an explicit meta-model we ensure the well-structuredness of these model instances and foster their preciseness as well as completeness.

\subsection{The 2-Dimensional Quality Meta-Model}

This model is based on the general idea of hierarchical models like FCM, \ie the breaking down of fuzzy criteria like \emph{learnability} into sub-criteria that are tangible enough to be assessed directly.
In contrast to other models, it introduces a rigorous separation of system \emph{properties} and \emph{activities} to be able to describe quality attributes and their impact on the usage of a software product precisely.

This approach is based on the finding that numerous criteria typically associated with usability, \eg \emph{learnability}, \emph{understandability}, and of course \emph{usability} itself, do not actually describe the properties of a system but rather the activities performed on (or with) the system.
It might be objected that these activities are merely expressed in the form of adjectives.
We argue, by contrast, that this leads precisely to the most prevalent difficulty of most existing quality models, namely to a dangerous mixture of activities and actual system properties.
A typical example of this problem can be found in~\cite{seffah:2006} where \emph{time behavior} and \emph{navigability} are presented as the same type of criteria.
Where \emph{navigability} clearly refers to the navigation activity carried out by the user of the system, \emph{time behavior} is a property of the system and not an activity.
One can imagine that this distinction becomes crucial, if the usability of a system is to be evaluated regarding different types of users:
The way a user navigates is surely influenced by the system, but is also determined by the individuality of the user.
In contrast, the response times of systems are absolutely independent of the user.
A simplified visualization of the system property and activity decompositions as well as their interrelations is shown in Fig.~\ref{fig:simple_model}.
The activities are based on Norman's action model~\cite{norman:1986}.
The whole model is described in detail in Sec.~\ref{sec:general_model}.

\insertfigure{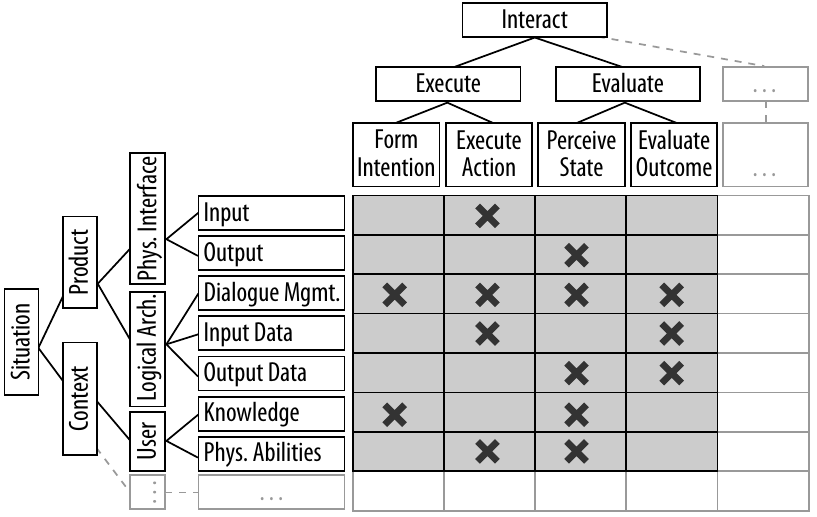}{Simplified quality model}
             {fig:simple_model}{width=0.8\linewidth}{ht}

The final goal of usability engineering is to improve the \emph{usage} of a system, \ie to create systems that support the activities that the user performs on the system.
Therefore, we claim that usability quality models must not only feature these activities as first-class citizens, but also precisely describe how properties of the system influence them and therewith ultimately determine the usability of the system.

\subsection{Facts, Activities, Attributes, \& Impacts}

Our usability model does not only describe the product, \ie the user interface, itself, but also comprises all relevant information about the situation of use (incl. the user).
To render this description more precisely the model distinguishes between \emph{facts} and \emph{attributes}.
Facts serve as a means to describe the situation of use in a hierarchical manner but do not contain quality criteria.
For example, they merely model that the fact \emph{user interface} consists of the sub-facts \emph{visual interface} and \emph{aural interface}.

\emph{Attributes} are used to equip the facts with desired or undesired low-level quality criteria like \emph{consistency}, \emph{ambiguousness}, or even the simple attribute \emph{existence}.
Thus, tuples of facts and attributes express system properties.
An example is the tuple \textsf{[Font Face$\vert${\scriptsize CONSISTENCY}]} that describes the consistent usage of font faces throughout the user interface.
Please note, that for clarity's sake the attributes are not shown in Fig.~\ref{fig:simple_model}.

The other part of the model consists of a hierarchical decomposition of the activities performed by a user as part of the interaction with the system.
Accordingly, the root node of this tree is the activity \emph{interact} that is subdivided into activities like \emph{execute} and \emph{evaluate} which in turn are broken down into more specific sub-activities.

Similar to facts, activities are equipped with attributes.
This allows us to distinguish between different properties of the activities and thereby fosters model preciseness.
Attributes typically used for activities are \emph{duration} and \emph{probability of error}.
The complete list of attributes is described in Sec.~\ref{sec:general_model}.

The combination of these three concepts enables us to pinpoint the impact that properties of the user interface (plus further aspects of the situation of use) have on the user interaction.
Here impacts are always expressed as a relation between fact-attribute-tuples and activity-attribute-tuples and qualified with the direction of the impact (positive or negative):

\begin{center}
\impact{Fact $\mathsf{f}$}{Attribute $\mathsf{a_1}$}
{Activity $\mathsf{a}$}{Attribute $\mathsf{a_2}$}{+/-}
\end{center}

For example, one would use the following impact description

\begin{center}
\impactn{Font Face}{Consistency}{Reading}{Duration}
\end{center}

to express that the consistent usage of font faces has a positive impact on the time needed to read the text.
Similarly the impact 

\begin{center}
\impactn{Input Validity Checks}{Existence}{Data Input}{Probability of Error}
\end{center}

is used to explain that the existence of validity checks for the input reduces the likelihood of an error.

\subsection{Tool Support}\label{tool-sup}

Our quality models are of substantial size (\eg the current model for maintainability has $>$ 800~model elements) due to the high level of detail.
We see this as a necessity and not a problem, since these models describe very complex circumstances.
However, we are well aware that models of this size can only be managed with proper tool support.
We have therefore developed a graphical editor, based on the \textsc{Eclipse} platform%
\footnote{\url{http://www.eclipse.org}}
that supports quality engineers in creating models and in adapting these models to changing quality needs by refactoring functionality.
Additionally, the editor provides quality checks on the quality models themselves, \eg it warns about facts that do not have an impact on any activity.

For the distribution of quality models the editor provides an export mechanism that facilitates exporting models (or parts thereof) to different target formats.
Supported formats are, \eg, simple graphs that illustrate the activity and system decomposition, but also full-fledged quality guideline documents that serve as the basis for quality reviews.
This export functionality can be extended via a plug-in interface.

\section{Usability Quality Model}
\label{sec:general_model}

Based on the critique of existing usability models described in Sec.~\ref{sec:related} and using the quality modeling approach based on the meta-model from Sec.~\ref{sec:qmm}, we propose a 2-dimensional quality model for usability.
The complete model is too large to be described in total, but we will highlight specific core parts of the model to show the main ideas.

Our approach to quality modeling includes \emph{high-level} and \emph{specific} models.
The aim of the high-level model is to define a basic set of facts, attributes, and activities that are independent of specific processes and domains.
It is simultaneously abstract and general enough to be reusable in various companies and for various products.
In order to fit to specific projects and situations the high-level models are refined and tailored into specific models.

\subsection{Goals}
\label{sec:goals}

In accordance with existing standards~\cite{iso9126-4:2004}, we see four basic principles needed for defining usability:

\begin{itemize}
\item \emph{Efficiency.} The utilization of resources.
\item \emph{Effectiveness.} The sharing of successful tasks.
\item \emph{Satisfaction.} The enjoyment of product use.
\item \emph{Safety.} The assurance of non-harmful behavior.
\end{itemize}

Fr{\o}kj{\ae}r, Hertzum, and Hornb{\ae}k~\cite{frokjar_2000_correlation} support the importance of these aspects:
``Unless domain specific studies suggest otherwise, effectiveness, efficiency, and satisfaction should be considered independent aspects of usability and all be included in usability testing.'' 
However, we do not use these principles directly for analysis, but rather to define the usability goals of the system.
The goals are split into several attributes of the activities inside the model.
For example, the effectiveness of the user interface depends on the probability of error for all activities of usage.
Therefore, all impacts on the attribute \emph{probability of error} of activities are impacts on the effectiveness and efficiency.
We describe more examples below after first presenting the most important facts, activity trees, and attributes.

\subsection{The Activity Subtree ``Interacting with the Product''}

The activity tree in the usability model has the root node \emph{use} that denotes any kind of usage of the software-based system under consideration.
It has two children, namely \emph{execution of secondary tasks} and \emph{interacting with the product}.
The former stands for all additional tasks a user has that are not directly related to the software product.
The latter is more interesting in our context because it describes the interaction with the software itself.
We provide a more detailed explanation of this subtree in the following.

\subsubsection{Activities.}
The activity \emph{interacting with the product} is further decomposed, based on the seven stages of action from Norman~\cite{norman:1986} that we arranged in a tree structure (Fig.~\ref{fig:activities}).
We believe that this decomposition is the key for a better understanding of the relationships in usability engineering.
Different system properties can have very different influences on different aspects of the use of the system.
Only if these are clearly separated will we be able to derive well-founded analyses.
The three activities, \emph{forming the goal}, \emph{executing}, and \emph{evaluating}, comprise the first layer of decomposition.
The first activity is the mental activity of deciding which goal the user wants to achieve.
The second activity refers to the actual action of planning and realizing the task.
Finally, the third activity stands for the gathering of information about the world's state and understanding the outcome.

\insertfigure{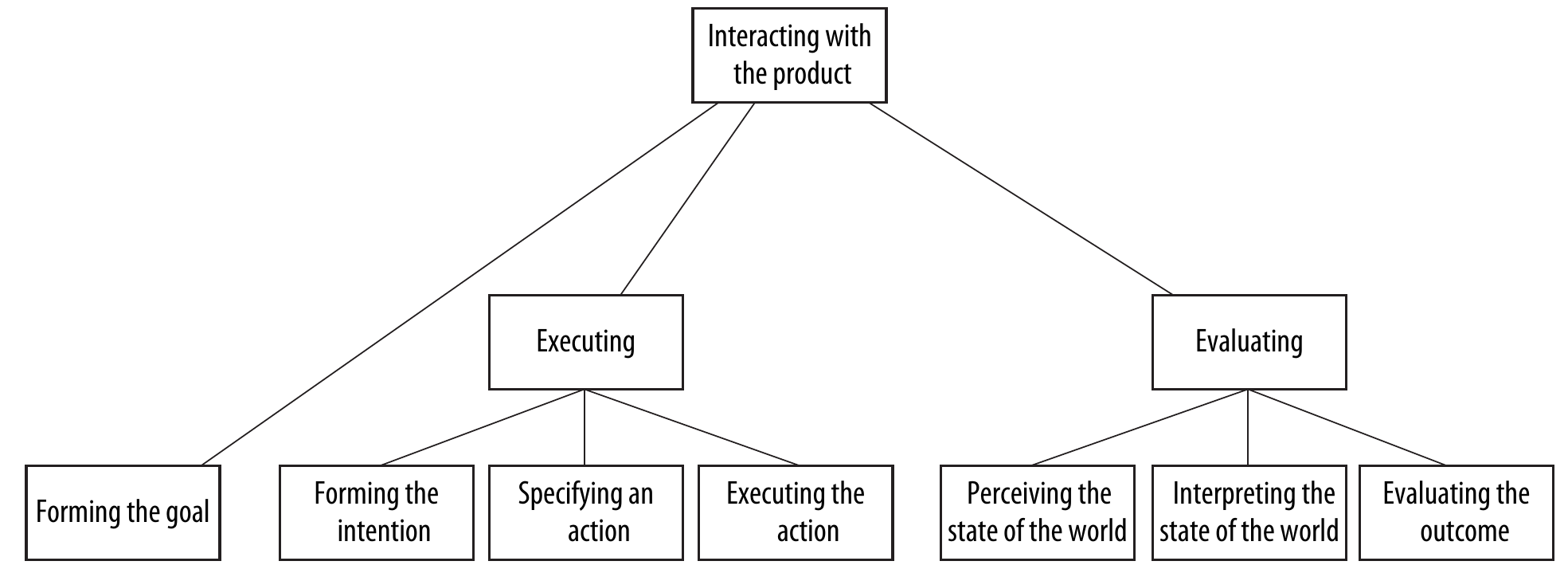}{The subtree for ``Interacting with the Product'' (adapted from~\cite{norman:1986})}
             {fig:activities}{width=\textwidth}{ht}

The \emph{executing} node has again three children:
First, the user forms his intention to do a specific action.
Secondly, the action is specified, \ie it is determined what is to be done.
Thirdly, the action is executed.
The \emph{evaluating} node is decomposed into three mental activities:
The user perceives the state of the world that exists after executing the action.
This observation is then interpreted by the user and, based on this, the outcome of the performed action is evaluated.
Scholars often use and adapt this model of action.
For example, Sutcliffe~\cite{sutcliffe:2002} linked error types to the different stages of action and Andre et~al.~\cite{andre:2001} developed the \textsc{User Action Framework} based on this model.

\subsubsection{Attributes.}

To be able to define the relation of the facts and activities to the general usability goals defined above, such as \emph{efficiency} or \emph{effectiveness}, we need to describe additional properties of the activities.
This is done by a simple set of attributes that is associated with the activities:

\begin{itemize}
\item \emph{Frequency.}
The number of occurrences of a task.

\item \emph{Duration.}
The amount of time a task requires.

\item \emph{Physical stress.}
The amount of physical requirements necessary to perform a task.

\item \emph{Cognitive load.}
The amount of mental requirements necessary to perform a task.
\item \emph{Probability of error.}
The distribution of successful and erroneous performances of a task.
\end{itemize}

As discussed in Sec.~\ref{sec:goals}, these activity attributes can be used to analyze the usability goals defined during requirements engineering.
We already argued that the effectiveness of a user interface is actually determined by the probability of error of the user tasks.
In our model, we can explicitly model which facts and situations have an impact on that.
The efficiency sets the frequency of an activity into relation to a type of resources:
time (duration), physical stress, or cognitive load.
We can explicitly model the impacts on the efficiency of these resources.
Further attributes can be used to assess other goals.

\subsection{The Fact Subtree ``Logical User Interface''}
\label{sec:logical_ui}

The fact tree in the usability model contains several areas that need to be considered in usability engineering, such as the physical user interface or the usage context.
By means of the \emph{user} component, important properties of the user can be described.
Together with the \emph{application} it forms the context of use.
The \emph{physical output devices} and the \emph{physical input devices} are assumed to be part of the physical user interface.
However, we concentrate on a part we consider very important:
the \emph{logical user interface}.
The decomposition follows mainly the logical architecture of a user interface as shown in Fig.~\ref{fig:arch}.

\insertfigure{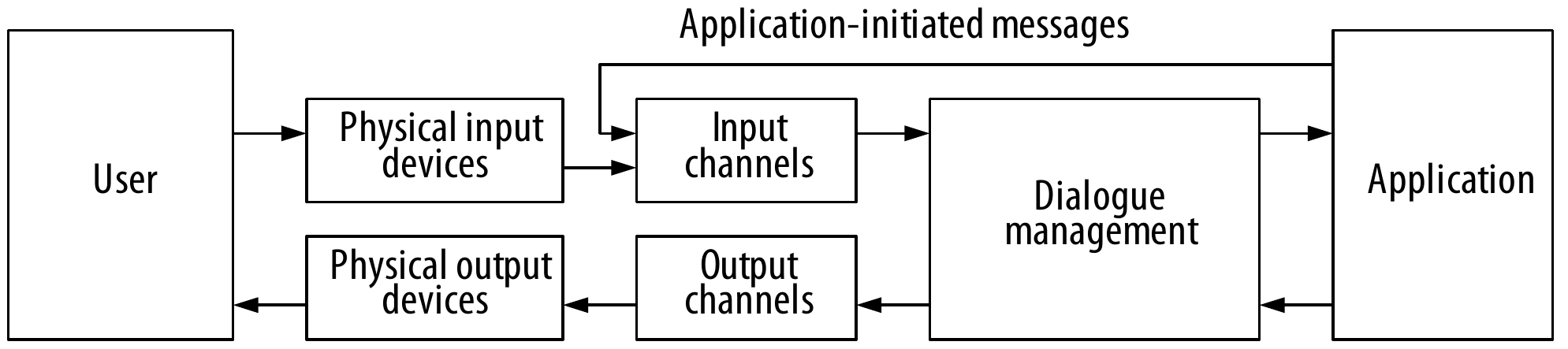}{The user interface architecture}{fig:arch}{width=120mm}{ht}

\subsubsection{Facts.}
The logical user interface contains \emph{input channels}, \emph{output channels}, and \emph{dialogue management}.
In addition to the architecture, we also add data that is sent via the channels explicitly:
\emph{input data} and \emph{output data}.
The architecture in Fig.~\ref{fig:arch} also contains a specialization of input data, \emph{application-initiated messages}.
These messages, which are sent by the \emph{application}, report interrupts of the environment or the application itself to the \emph{dialogue management} outside the normal response to inputs.

\subsubsection{Attributes.}

The attributes play an important role in the quality model because they are the properties of the facts that can actually be assessed manually or automatically.
It is interesting to note that it is a rather small set of attributes that is capable of describing the important properties of the facts.
These attributes are also one main building block that can be reused in company- or domain-specific usability models.
Moreover, we observe that the attributes used in the usability model differ only slightly from the ones contained in the maintainability model of~\cite{deissenb:2006}.
Hence, there seems to be a common basic set of those attributes that is sufficient -- in combination with facts -- for quality modeling.

\begin{itemize}

\item \emph{Existence}.
The most basic attribute that we use is whether a fact exists or not.
The pure existence of a fact can have a positive or negative impact on some activities.

\item \emph{Relevance}.
When a fact is relevant, it means that it is appropriate and important in the context in which it is described.

\item \emph{Unambiguousness.}
An unambiguous fact is precise and clear.
This is often important for information or user interface elements that need to be clearly interpreted.

\item \emph{Simplicity}.
For various facts it is important that in some contexts they are simple.
This often means something similar to small and straightforward.

\item \emph{Conformity}.
There are two kinds of conformity:
conformity to existing standards and guidelines, and conformity to the expectations of the user.
In both cases the fact conforms to something else, \ie it respects and follows the rules or models that exist.

\item \emph{Consistency}.
There are also two kinds of consistency:
internal consistency and external consistency.
The internal consistency means that the entire product follows the same rules and logic.
The external consistency aims at correspondence with external facts, such as analogies, or a common understanding of things.
In both cases it describes a kind of homogeneous behavior.

\item \emph{Controllability}.
A controllable fact is a fact which relates to behavior that can be strongly influenced by the actions of the user.
The user can control its behavior.

\item \emph{Customizability}.
A customizable fact is similar to a controllable fact in the sense that the user can change it.
However, a customizable fact can be preset and fixed to the needs and preferences of the user.

\item \emph{Guardedness}.
In contrast to customizability and controllability, a guarded fact cannot be adjusted by the user.
This is a desirable property for some critical parts of the system.

\item \emph{Adaptability}.
An adaptive fact is able to adjust to the user's needs or to its context dependent on the context information.
The main difference to customizability is that an adaptive fact functions without the explicit input of the user.
\end{itemize}

\subsection{Examples}

The entire model is composed of the activities with attributes, the facts with the corresponding attributes and the impacts between attributed facts and attributed activities.
The model with all these details is too large to be described in detail, but we present some interesting examples:
triplets of an attributed fact, an attributed activity and a corresponding impact.
These examples aim to demonstrate the structuring that can be achieved by using the quality meta-model as described in Sec.~\ref{sec:qmm}.

\paragraph{Consistent Dialogue Management.}
A central component in the logical user interface concept proposed in Sec.~\ref{sec:logical_ui} is the \emph{dialogue management}.
It controls the dynamic exchange of information between the product and the user.
In the activities tree, the important activity is carried out by the user by interpreting the information given by the user interface.
One attribute of the dialogue management that has an impact on the interpretation is its \emph{internal consistency}.
This means that its usage concepts are similar in the entire dialogue management component. 
The corresponding impact description:

\begin{center}
\impactn{Dialogue Management}{Internal Consistency}{Interpretation}{Prob. of Error}
\end{center}

Obviously, this is still too abstract to be easily assessed.
This is the point where company-specific usability models come in.
This general relationship needs to be refined for the specific context.
For example, menus in a graphical user interface should always open the same way.

\paragraph{Guarded Physical Interface.}
The usability model does not only contain the logical user interface concept, but also the physical user interface.
The \emph{physical interface} refers to all the hardware parts that the user interacts with in order to communicate with the software-based system.
One important attribute of such a physical interface is \emph{guardedness}.
This means that the parts of the interface must be guarded against unintentional activation. 
Hence, the guardedness of a physical interface has a positive impact on the \emph{executing} activity:

\begin{center}
\impactn{Physical Interface}{guardedness}{Executing}{Probability of Error}
\end{center}

A physical interface that is not often guarded is the touchpad of a notebook computer.
Due to its nearness to the location of the hands while typing, the cursor might move unintentionally.
Therefore, a usability model of a notebook computer should contain the triplet that describes the impact of whether the touchpad is guarded against unintentional operation or not.

\section{Case Study: Modeling the ISO~15005}
\label{sec:case_study}

To evaluate our usability modeling approach we refine the high-level model described in Sec.~\ref{sec:general_model} into a specific model based on the ISO~15005~\cite{iso15005:2002}.
This standard describes ergonomic principles for the design of \emph{transport information and control systems} (TICS).
Examples for TICS are driver information systems (\eg navigation systems) and driver assistance systems (\eg cruise control).
In particular, principles related to dialogues are provided, since the design of TICS must take into consideration that a TICS is used in addition to the driving activity itself.

The standard describes three main \emph{principles} which are further subdivided into eight \emph{sub-principles}.
Each sub-principle is motivated and consists of a number of \emph{requirements} and/or \emph{recommendations}.
For each requirement or recommendation a number of examples are given.

For example, the main principle \emph{suitability for use while driving} is decomposed among others into the sub-principle \emph{simplicity}, \ie the need to limit the amount of information to the task-dependent minimum.
This sub-principle consists, among others, of the recommendation to optimize the driver's mental and physical effort.
All in all the standard consists of 13~requirements, 16~recommendations, and 80~examples.

\subsection{Approach}

We follow two goals when applying our method to the standard:
First, we want to prove that our high-level usability model can be refined to model such principles.
Secondly, we want to discover inconsistencies, ill-structuredness, and implicitness of important information.

Our approach models every element of the standard (\eg high-level principles, requirements, etc.) by refinement of the high-level model.
For this, the meta-model elements (\eg facts, attributes, impacts, etc.) are used.
We develop the specific model by means of the tool described in Sec.~\ref{tool-sup}.
The final specific model consists of 41~facts, 12~activities, 15~attributes, 48~attributed facts, and 51~impacts.

\subsection{Examples}

To illustrate how the elements of the standard are represented in our specific model, we present the following examples.

\paragraph{Representation of Output Data.}
An element in the logical user interface concept proposed in Sec.~\ref{sec:logical_ui} is the \emph{output data}, \ie the information sent to the driver.
A central aspect is the representation of the data.
One attribute of the representation that has an impact on the interpretation of the state of the system is its \emph{unambiguousness}, \ie that the representation is precise and clear.
This is especially important so that the driver can identify the exact priority of the data.
For example, warning messages are represented in a way that they are clearly distinguishable from status messages.

\begin{center}
\impactn{Output Data}{Unambiguousness}{Interpretation}{Probability of Error}
\end{center}

Another attribute of the representation that has an impact on the interpretation is the \emph{internal consistency}.
If the representations of the output data follow the same rules and logic, it is easier for the driver to create a mental model of the system.
The ease of creating a mental model has a strong impact on the ease of interpreting the state of the system:

\begin{center}
\impactn{Output Data}{Internal Consistency}{Interpretation}{Duration}
\end{center}

One attribute of the representation that has an impact on the perception is \emph{simplicity}.
It is important for the representation to be simple, since this makes it easier for the driver to perceive the information:

\begin{center}
\impactn{Output Data}{Simplicity}{Perception}{Cognitive Load}
\end{center}

\paragraph{Guarded Feature.}

A TICS consists of several features which must not be used while driving the vehicle.
This is determined by the manufacturer as well as by regulations.
One important attribute of such features is its \emph{guardedness}.
This means that the feature is inoperable while the vehicle is moving.
This protects the driver from becoming distracted while using the feature.
The guardedness of certain features has a positive impact on the \emph{driving} activity:

\begin{center}
\impactn{Television}{guardedness}{Driving}{Probability of Error}
\end{center}

\subsection{Observations \& Improvements}

As a result of the meta-model-based analysis, we found the following inconsistencies and omissions:

\paragraph{Inconsistent Main Principles.}
One of the three main principles, namely \emph{suitability for the driver}, does not describe any activity. 
The other two principles use the activities to define the high-level usability goals of the system.
For example, one important high-level goal is that the TICS dialogues do not interfere with the driving activity.
Hence, we suggest that every main principle should describe an activity and the high-level goals of usability should be defined by means of the attributes of the user's activities.

\paragraph{Mixed Sub-Principles.}
The aspects described by the sub-principles are mixed:
Three sub-principles describe activities without impacts, three describe facts without impacts, and the remaining two describe impacts of attributes on activities.
This mix-up of the aspects described by the sub-principles must be resolved. 

We believe that in order to make a design decision it is crucial for the software engineer to know which high-level goals will be influenced by it.
Sub-principles which only describe attributes of system entities do not contribute toward design decisions.
The same holds true for sub-principles which only describe activities, since they are not related to system entities.
For this reason we suggest that all sub-principles that only describe activities should be situated at the main principle level, while those sub-principles that describe software entities should be situated at the requirement level.

\paragraph{Requirements with Implicit Impacts.}
9 out of 13~requirements do not explicitly describe impacts on activities.
Requirements serve to define the properties which the system entities should fulfill.
If a requirement does not explicitly describe its impacts on activities, the impact could be misunderstood by the software engineer.
Hence, we suggest that requirements should be described by attributed facts and their impacts on activities.

\paragraph{Incomplete Examples.}
14 out of 80~examples only describe facts and their attributes, leaving the impacts and activities implicit.
To provide complete examples we suggest that the examples should be described with explicit impacts and activities.

\section{Discussion}
\label{sec:discussion}

The usability model acts as a central knowledge base for the usability-related relationships in the product and process.
It documents in a structured manner how the properties of the system, team, and organization influence different usage activities.
Therefore, it is a well-suited basis for quality assurance (QA).
It can be used in several ways for constructive as well as analytical QA.
Some of these have been shown to be useful in an industrial context
w.r.t.~maintainability models.

\paragraph{Constructive QA.}
The knowledge documented in the quality model aids all developers and designers in acquiring a common understanding of the domain, techniques, and influences.
This common understanding helps to avoid misunderstandings, and improvements to the quality model become part of a continuous learning process for all developers.
For example, by describing the properties of the system artifacts, a glossary or terminology is built and can be easily generated into a document.
This glossary is a living artifact of the development process, not only because it is a materiality itself, but also because it is inside and part of a structured model.
Hence, by learning and improving the way developers work, it is possible to avoid the introduction of usability defects into the product.

\paragraph{Analytical QA.}
The identified relationships in the usability model can also be used for analytical QA.
With our quality model we aim to break down the properties and attributes to a level where we can measure them and, therefore, are easily able to give concrete instructions in analytical QA.
In particular, we are able to generate guidelines and checklists for reviews from the model.
The properties and attributes are there and subsets can easily be selected and exported in different formats so that developers and reviewers always have the appropriate guidelines at hand.
Moreover, we annotate the attributed properties in the model, whether they are automatically, semi-automatically, or only manually assessable.
Hence, we can identify quality aspects that can be analyzed automatically straightforwardly.
Thus, we are able to use all potential benefits of automation.

\paragraph{Analyses and Predictions.}
Finally, more general analysis and predictions are possible based on the quality model.
One reason to organize the properties and activities in a tree structure is to be able to aggregate analysis to higher levels.
This is important to get concise information about the quality of the system.
To be able to do this, the impacts of properties on activities must be quantified.
For example, the usability model is a suitable basis for cost/benefit analysis because the identified relationships can be quantified and set into relation to costs similar to the model in~\cite{wagner:issta06}.
In summary, we are able to aid analytical QA in several ways by utilizing the knowledge coded into the model.

\section{Conclusion}
\label{sec:conclusions}

Usability is a key criteria in the quality of software systems, especially for its user.
It can be decisive for its success on the market.
However, the notion of usability and its measurement and analysis are still not fully understood.
Although there have been interesting advances by consolidated models, e.g.~\cite{seffah:2006}, these models suffer from various shortcomings, such as inconsistencies in the dimensions used.
An approach based on an explicit meta-model has proven to be useful for the quality attribute \emph{maintainability}.
Hence, we propose a comprehensive usability model that is based on the same meta-model.

Using the meta-model and constructing such a usability model allows us to describe completely the usability of a system by its facts and their relationship with (or impact on) the activities of the user.
We support the consistent and unambiguous compilation of the usability knowledge available.
The general model still needs to be refined for specific contexts that cannot be included in a general model.
By utilizing a specific usability model, we have several benefits, such as the ability to generate guidelines and glossaries or to derive analyses and predictions.

The usefulness of this approach is demonstrated by a case study in which an ISO standard is modeled and several omissions are identified.
For example, the standard contains three sub-principles which describe activities, but no impacts on them, as well as nine requirements that have no described impacts.
This hampers the justification of the guideline:
A rule that is not explicitly justified will not be followed.

For future work we plan to improve further the general usability model and to carry out more case studies in order to validate further the findings of our current research.
Furthermore, other quality attributes, \eg \emph{reliability}, will also be modeled by means of the meta-model to investigate whether this approach works for all attributes.
If this be the case, the different models can be combined, since they are all based on a common meta-model.

\bibliography{dsvis07}

\noindent \textcopyright Springer-Verlag. The final publication is available at \\ http://link.springer.com/chapter/10.1007/978-3-540-92698-6\_7.

\end{document}